\documentclass[12pt]{article} 
\pdfoutput=1
     \usepackage{amsmath}
  \usepackage{graphicx}
  \usepackage{color}
  \usepackage{subfigure}
  \usepackage{tikz}
  \newlength{\abstractwidth}
 \usepackage[numbers,sort&compress]{natbib}
 \usepackage{amsmath,amssymb}
\usepackage{graphicx}
\usepackage{caption}
\usepackage{lipsum}
\usepackage[utf8]{inputenc}
  \usepackage{mciteplus} 

\usepackage[colorlinks = true,
            linkcolor = blue,
            urlcolor  = blue,
            citecolor = blue,
            anchorcolor = blue]{hyperref}

\hypersetup{colorlinks=true,linkcolor=blue}

\usepackage[title]{appendix}
\usepackage{comment}
\usepackage{braket}

\def\XXint#1#2#3{{\setbox0=\hbox{$#1{#2#3}{\int}$}
     \vcenter{\hbox{$#2#3$}}\kern-.5\wd0}}

\makeatletter
\def\blfootnote{\xdef\@thefnmark{}\@footnotetext}
\makeatother

  \newcommand{\be}{\begin{equation}}
  \newcommand{\bea}{\begin{eqnarray}}
  \newcommand{\eea}{\end{eqnarray}}
  \newcommand{\beq}{\begin{equation}}
  \newcommand{\ee}{\end{equation}}
  \newcommand{\eeq}{\end{equation}}

  \def\ba{\begin{eqnarray}}
  \def\ea{\end{eqnarray}}

 \def\simleq{\; \raise0.3ex\hbox{$<$\kern-0.75em
      \raise-1.1ex\hbox{$\sim$}}\; }
 \def\simgeq{\; \raise0.3ex\hbox{$>$\kern-0.75em
      \raise-1.1ex\hbox{$\sim$}}\; }

\begin{document}

\begin{titlepage}
  \bigskip

  \bigskip\bigskip

  \bigskip

\begin{center}
{\Large \bf{}}
 \bigskip
{\Large \bf {Pulling Out the Island with Modular Flow}} 
\bigskip
\bigskip
   \bigskip
\bigskip
\end{center}

  \begin{center}

 \bf {Yiming Chen}
  \bigskip \rm
\bigskip
 
 \rm

Jadwin Hall, Princeton University,  Princeton, NJ 08540, USA\\

  \end{center}

 \bigskip\bigskip

\begin{abstract}

Recent works have suggested that the entanglement wedge of Hawking radiation coming from an AdS black hole will include an island inside the black hole interior after the Page time. In this paper, we propose a concrete way to extract the information from the island by acting only on the radiation degrees of freedom, building on the equivalence between the boundary and bulk modular flow. We consider examples with black holes in JT gravity coupled to baths. In the case that the bulk conformal fields contain free massless fermion field, we provide explicit bulk picture of the information extraction process, where we find that one can almost pull out an operator from the island to the bath with modular flow.

 \end{abstract}
\bigskip \bigskip \bigskip
\blfootnote{ymchen.phys@gmail.com}

  \end{titlepage}

   \tableofcontents


\section{Introduction}

Recent works \cite{Penington:2019npb,Almheiri:2019psf,Almheiri:2019hni} have suggested that the Page curve \cite{Page:1993wv,Page:2013dx}, which is a characteristic result of unitary black hole evaporation, can be calculated simply within semiclassical gravity, as long as one uses the correct formula for computing entropies in a gravitational system  (see also recent related works \cite{Akers:2019wxj,Fu:2019oyc,Akers:2019nfi,Almheiri:2019yqk,Rozali:2019day,Bousso:2019ykv,Almheiri:2019psy,Zhao:2019nxk}). The authors in \cite{Penington:2019npb,Almheiri:2019psf,Almheiri:2019hni} considered models with an AdS black hole coupled to a bath which absorbs the radiation from the black hole, where the whole system has a holographic dual as a boundary quantum mechanical system coupled to a bath. As shown in \cite{Penington:2019npb,Almheiri:2019psf}, by applying the standard Ryu-Takayanagi (RT) formula \cite{Ryu:2006bv} with its generalization including time dependence \cite{Hubeny:2007xt} and quantum correction \cite{Faulkner:2013ana,Engelhardt:2014gca}, one finds that the minimal quantum extremal surface has a phase transition at the Page time, which then gives the entropy curve for the black hole in accordance with unitarity. However, the traditional way of calculating the entropy for the radiation, which is Hawking's calculation \cite{Hawking:1974sw}, would still lead to result that contradicts with unitarity. As hinted in \cite{Penington:2019npb,Almheiri:2019psf}, a new way to compute the entropy for the radiation is needed. 

In \cite{Almheiri:2019hni}, a ``doubly holographic" set-up led the authors to put forth a new prescription for computing the fine-grained von Neumann entropy of the Hawking radiation, which involves taking into account the existence of the entanglement ``island":
\begin{equation}\label{newentropyrule}
    S[\boldsymbol{\rho(\textrm{\textbf{rad}})}] = \textrm{min} \left\{  \textrm{ext}_{I} \left[ S[\rho (\textrm{rad}\cup I)]  + \frac{\textrm{Area}[\partial I]}{4G_N}\right] \right\}.
\end{equation}
The prescription instructs us to search for all possible islands that can extremize the generalized entropy functional for the union of the radiation and the island, and then look for the minimal one. Here we adopted the convention from \cite{Almheiri:2019yqk} that the bold symbols or texts such as ``\textbf{rad}" refer to the full non-perturbative gravity description, or the dual quantum mechanical description, while ordinary texts like ``rad" refer to the description within semiclassical gravity, and ``rad" stands for radiation. We will call the density matrix $\boldsymbol{\rho}$ as the fine-grained density matrix, which is calculated by tracing out other degrees of freedom in the full non-perturbative gravity description, or in the dual quantum mechanical model, while $\rho$ is the density matrix calculated from the semiclassical bulk physics.
With this formula, one finds that the entropy curve of the radiation also agress with unitarity, thus a black hole information paradox is prevented. Direct derivations of this formula from replica calculations are presented in \cite{Almheiri:2019qdq,Penington:2019kki}.

An important consequence of this formula is that the island is included in the entanglement wedge of the radiation. In the black hole evaporation set-up in \cite{Penington:2019npb,Almheiri:2019psf,Almheiri:2019hni}, the island is inside the black hole interior, and entanglement wedge reconstruction \cite{Czech:2012bh,Headrick:2014cta,Dong:2016eik} suggests that part of the black hole interior is secretly encoded in the Hawking radiation. Thus one knows that although the island is isolated and far away from the radiation, by acting only on the radiation degrees of freedom in a sufficiently complicated way, one should in principle be able to extract information from the island to the radiation. However, a concrete proposal of how to do it is lacking (though see the note added in the end of this section), and it is not clear whether the information extraction can be done while maintaining the semiclassical bulk picture. In this paper, we give an affirmative answer to this question by proposing a concrete way to pull out information from the island. 

Our result is built on the established role of the modular Hamiltonian and modular flow in entanglement wedge reconstruction. In the AdS/CFT correspondence, the modular Hamiltonian of a boundary region of the CFT is dual to the area operator on the RT surface, plus the bulk modular Hamiltonian of bulk quantum fields in the entanglement wedge, with higher order corrections in the $G_N$ expansion \cite{Jafferis:2014lza,Jafferis:2015del} (see also \cite{Harlow:2016vwg} from a quantum error correction perspective). This is further strengthened to an equality in the case of quantum extremal surfaces \cite{Dong:2017xht} (see also recent discussion in \cite{Dong:2019piw}). One consequence of such a duality is that to the leading order of bulk perturbation theory, boundary modular flow is equivalent to the bulk modular flow \cite{Faulkner:2017vdd}, where modular flow is defined as an unitary evolution using the modular Hamiltonian. The concept of modular flow was found to be useful in the context of entanglement wedge reconstruction \cite{Faulkner:2017vdd,Almheiri:2017fbd} and bulk reconstruction \cite{Chen:2018rgz,Faulkner:2018faa}. 

Since the dual description of the modular Hamiltonian closely follows from the RT formula, the new entropy formula (\ref{newentropyrule}) naturally gives rise to a generalized formula in the cases with islands:
\begin{equation}
    \boldsymbol{K  [\textbf{rad}]} = \frac{\widehat{\textrm{Area}}[\partial I]}{4G_N} + K[\textrm{rad}\cup I].
\end{equation}
In the formula, the hat on ``Area" means that we should treat it as an operator in the semiclassically quantized bulk theory \cite{Jafferis:2014lza}. $\boldsymbol{K  [\textbf{rad}]}$ refers to the microscopic modular Hamiltonian of the radiation in the exact non-perturbative description, while $K[\textrm{rad}\cup I]$ refers to the bulk modular Hamiltonian of the quantum fields in the semiclassical description on the union of the radiation and the island. Now, if one applies a modular flow with the microscopic modular Hamiltonian on the radiation degrees of freedom, in the gravity picture it will correspond to a modular flow on the union of the radiation and the island. Due to the entanglement between the quantum fields in the radiation and the island, which is necessary for the existence of the island, the bulk modular Hamiltonian is a non-local operator which couples the radiation and the island directly. Our goal is to show that one can utilize this inherent nonlocalness to pull out information from the island.
To simplify our discussion, instead of studying the evaporating black hole examples, we will instead set the stage on the simpler set-ups discussed in \cite{Almheiri:2019yqk}, where the black holes are in equilibrium with the baths, while one can still formulate an information paradox and islands still play important roles in resolving the puzzle. The lesson of our discussion applies to general situations where islands show up.

The paper is organized as follows. In section \ref{extremal}, we first discuss the example with an extremal black hole coupled to a bath. In sec. \ref{general}, we provide general arguments on how modular flow pulls out the information from the island. In sec. \ref{sec:formulafermion} and sec. \ref{sec:pullout}, we consider an example that the bulk quantum fields contain a free massless fermion field, where by utilizing the exact result of modular Hamiltonian on disjoint intervals derived in \cite{Casini:2009vk}, we are able to show how the proposal works explicitly. In sec. \ref{twoside}, we generalize the discussion to the example with non-extremal black holes coupled to baths, and we find that one can almost extract information from the black hole interior perfectly at late time.

~

Note added: after the completion of this work, a different idea of extracting information from the island using the Petz map \cite{Cotler:2017erl,Chen:2019gbt} was also proposed in \cite{Penington:2019kki}. It would be interesting to understand its relation with our proposal.

\section{Extremal black hole coupled to a bath}\label{extremal}

\subsection{General argument}\label{general}

\begin{figure}[t!]
\centering  

\begin{tikzpicture}[thick,scale = 1]

\fill[green, opacity = 0.05] (0,-2.5) -- (-2,-2.5) -- (-2,2.5) -- (0,2.5); 

 \draw (0,-2.5) -- (0,2.5);
 \draw (-2,-2.5) -- (-2,2.5);
 \draw (-2,0) -- (0,2);
 \draw (-2,0) -- (0,-2);
 \draw (2,0) -- (0,2);
 \draw (2,0) -- (0,-2);
 \draw (0.8,0.5) node{Bath};
 \draw (-0.8,0.5) node{AdS$_2$};
  \draw (0,-2.8) node{$x=0$};
 \draw[->] (2.5,1) -- (3.5,1) node[right]{$x$};
 \draw[->] (2.5,1) -- (2.5,2) node[above]{$t$};
 \draw[->] (2.5,1) -- (3.2,1.7) node[right]{$u_+$};
 \draw[->] (2.5,1) -- (1.8,1.7) node[left]{$u_-$};

 
 \draw (5,0) node{$\equiv$};
 \filldraw[draw = black] (6,0) circle (0.2);
 \draw (6,0.5) node{QM};
 \draw (6,0) -- (9,0);
  \draw (7.5,0.5) node{CFT};
\end{tikzpicture}

\caption{Left: the Penrose diagram of the system. Dynamical gravity only lives in the green region. Right: the quantum mechanical description of the system, which involves a (0+1) dimensional system coupled to a CFT on a half infinite line.}
\label{fig:penrose0}
\end{figure}
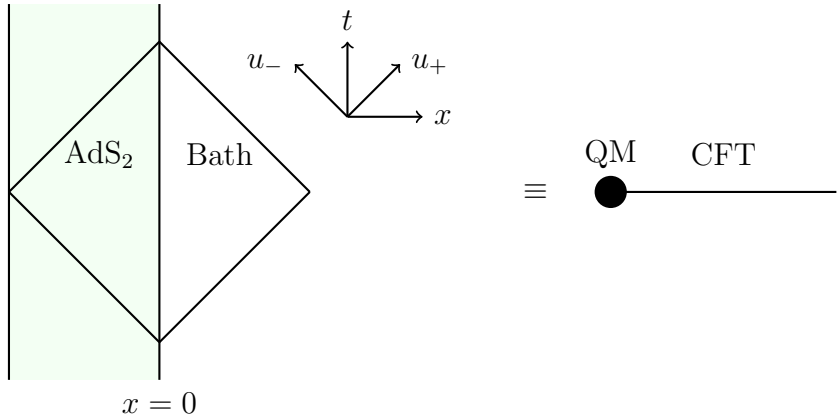

Before discussion of the modular flow, we first review the set up of the extremal black hole example discussed in \cite{Almheiri:2019yqk}. More specifically, we consider a zero temperature black hole in the two-dimensional Jackiw-Teitelboim gravity \cite{Jackiw:1984je,Teitelboim:1983ux,Almheiri:2014cka} coupled to two-dimensional CFT with central charge $c \gg 1$.\footnote{We are not requiring the conformal field theory to have a holographic dual.} On the boundary of the AdS$_2$ spacetime we pick the transparent boundary condition, and let the CFT to continue into a bath smoothly, where there is no gravity and the metric is fixed to be flat. The metric in the AdS$_2$ region is
\begin{equation}
    ds^2 = \frac{-dt^2 + dx^2}{x^2}, \quad x <0,
\end{equation}
and the dilaton profile in the AdS$_2$ region is
\begin{equation}
    \phi = \phi_0 - \frac{\phi_r}{x},
\end{equation}
where $\phi_0$ corresponds to the extremal entropy of the black hole. When the conformal fields are in the Poincar\'{e} vacuum, one can choose a special gauge and write the fixed metric of the bath as
\begin{equation}
    ds^2 = -dt^2 + dx^2, \quad x>0.
\end{equation}
At $x=0$ where the AdS$_2$ and the bath are joined together, one picks transparent boundary conditions for the bulk fields. The Penrose diagram of the system is shown in fig. \ref{fig:penrose0}. The conformal fields are in the Poincar\'{e} vacuum, and the stress tensor vanishes everywhere (apart from the piece coming from the conformal anomaly). Equivalently, one can view the diamond-shaped region outside the horizon in AdS$_2$ and the bath as conformally equivalent to the Minkowski spacetime, and the conformal fields are in the Minkowski vacuum with respect to the global time coordinate $t$. 

We will assume that gravitational part of the system has an dual description as a $(0+1)$-d quantum mechanical system. If there isn't such a dual, the idea of the discussion here should carry over to situations where a dual theory is known. In the dual picture, we have a quantum mechanical system coupled to a half infinite line where the CFT lives (see the right figure of fig. \ref{fig:penrose0}), and the combined system is put in the ground state.

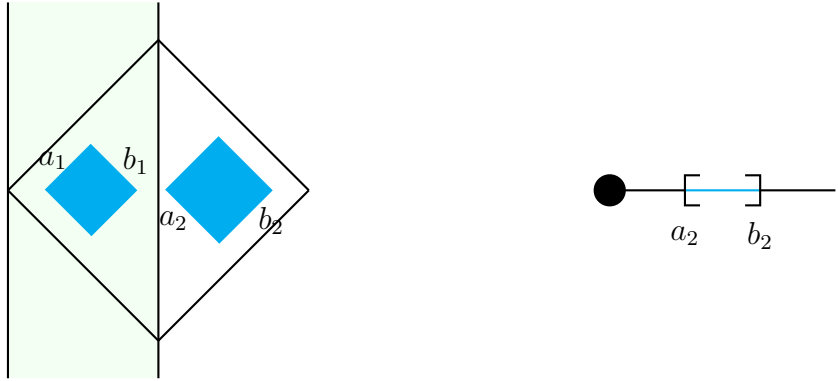
\begin{figure}[t!]
\centering

 \begin{tikzpicture}[thick,scale = 1]
 
 \fill[green, opacity = 0.05] (0,-2.5) -- (-2,-2.5) -- (-2,2.5) -- (0,2.5); 
 
 \draw (0,-2.5) -- (0,2.5);
 \draw (-2,-2.5) -- (-2,2.5);
 \draw (-2,0) -- (0,2);
 \draw (-2,0) -- (0,-2);
 \draw (2,0) -- (0,2);
 \draw (2,0) -- (0,-2);

  \draw (0.2,-0.4) node{$a_2$};
  \draw (1.5,-0.4) node{$b_2$};
  \draw (-0.3,0.4) node{$b_1$};
  \draw (-1.4,0.4) node{$a_1$};
\filldraw[cyan] (0.1,0) -- (0.8,0.7)--(1.5,0) -- (0.8,-0.7);
\filldraw[cyan] (-1.5,0) -- (-0.9,0.6)--(-0.3,0) -- (-0.9,-0.6);
 \filldraw[draw = black] (6,0) circle (0.2);
 \draw (6,0) -- (7,0);
 \draw[cyan, thick] (7,0) -- (8,0);
 \draw (8,0) -- (9,0);
  \draw[thick] (7.2,-0.2)--(7,-0.2) -- (7,0.2) -- (7.2,0.2);
  \draw[thick] (7.8,-0.2)--(8,-0.2) -- (8,0.2) -- (7.8,0.2);
  
  \draw (7,-0.6) node{$a_2$};
  \draw (8,-0.6) node{$b_2$};
\end{tikzpicture}

\caption{The entanglement wedge of region $\boldsymbol{[a_2 ,b_2]}$ in the quantum mechanical description contains the region $[a_2 ,b_2]$ itself plus an island $[a_1 ,b_1]$ (the blue regions in the left figure). }
\label{fig:penrose1}
\end{figure}

As instructed by (\ref{newentropyrule}), if one wants to calculate the microscopic von Neumann entropy of a region $\boldsymbol{[a_2,b_2]}$ in the bath, then in the gravity calculation, one should take into account of a possible entanglement island outside the horizon. One occasion such an island arises is when\footnote{We work in units such that $4G_N = 1$ as in \cite{Almheiri:2019yqk}.}
\begin{equation}\label{parameter}
   a_2 \ll \frac{\phi_r}{c} \ll b_2,\quad \log \left(\frac{cb_2}{\phi_r}\right) > \frac{12\phi_0}{c} + \mathcal{O}(1).
\end{equation}
Under this parameter region, the entanglement wedge of $\boldsymbol{[a_2 ,b_2]}$ contains the region $[a_2 ,b_2]$ itself plus an island $[a_1 ,b_1]$ (see fig. \ref{fig:penrose1}), with
\begin{equation}\label{parameter2}
    a_{1} \approx -b_2, \quad b_{1} \approx - \frac{6\phi_r}{c}.
\end{equation}
With (\ref{newentropyrule}), the von Neumann entropy of the region $\boldsymbol{[a_2 , b_2]}$ in the microscopic theory can be calculated via
\begin{equation}
    S(\boldsymbol{[a_2 , b_2]}) = 2 \phi_0 - \frac{\phi_r}{a_1} - \frac{\phi_r}{b_1} + S_{\textrm{bulk}} ([a_1,b_1] \cup [a_2 ,b_2]),
\end{equation}
where $S_{\textrm{bulk}} ([a_1,b_1] \cup [a_2 ,b_2])$ is the von Neumann entropy of the bulk quantum fields on $[a_1,b_1] \cup [a_2 ,b_2]$, calculated in the semiclassical gravity picture.

It is already surprising that semiclassical gravity knows about the fine-grained entropy. However, once this is established, much more would follow from the entropy formula. In particular, the modular Hamiltonian of the region $\boldsymbol{[a_2,b_2]}$ in the quantum mechanical description, which we denote as $\boldsymbol{K([a_2 , b_2])}$ or simply $\boldsymbol{K}$, has a bulk expression as the area operator plus the bulk modular Hamiltonian of the quantum fields on the region $[a_1,b_1] \cup [a_2,b_2]$:
\begin{equation}\label{JLMS}
    \boldsymbol{K([a_2 , b_2])} = \hat{\phi} (a_1) + \hat{\phi} (b_1) + K([a_1 , b_1] \cup [a_2,b_2]) + \mathcal{O}\left(1/c\right).
\end{equation}
In the formula, we put the hats on $\phi$ to stress that we should view them as operators in the semiclassically quantized bulk theory. The equation (\ref{JLMS}) should be valid within the code subspace, that is when we consider variations around the vacuum state such that the locations $a_1,b_1$ of the quantum extremal surfaces  stay approximately unchanged. In the current setup, since the locations of the quantum extremal surfaces have a non-trivial dependence on the central charge $c$ as in (\ref{parameter2}), we should only apply the equality (\ref{JLMS}) to situations where the bulk state is perturbed by only an order one number of light operators in the CFT, and keep in mind possible corrections of order $1/c$ in the formula as written.\footnote{We thank Juan Maldacena for clarification on this point.}

One consequence of (\ref{JLMS}) is that for a bulk operator $\varphi(x)$ inside the entanglement wedge (i.e. $x \in [a_1 , b_1]\cup [a_2,b_2]$),\footnote{Here we've implicitly assumed that the gravitational dressing is done properly. One can refer to \cite{Almheiri:2018xdw} for discussion of gravitational dressing in JT gravity.} one has
\begin{equation}\label{commutator}
    [ \boldsymbol{K([a_2 , b_2])} , \varphi(x)] =  [ K([a_1 , b_1] \cup [a_2,b_2]) , \varphi(x)] + \mathcal{O}\left(1/c\right).
\end{equation}
At this point, we would like to stress again the difference between the boldface $\boldsymbol{K([a_2 , b_2])}$ and the semiclassical modular Hamiltonian $K([a_2,b_2])$. Although from the quantum mechanical system point of view, both are operators supported on the same region, the latter one only knows about the semiclassical dynamics in the causal diamond of $[a_2,b_2]$, while the former one secretly knows about the information of the island via the non-perturbative description. Since in most of the following discussion we will not use the coarse-grained modular Hamiltonian $K([a_2,b_2])$, we will use $K$ to denote $K([a_1 , b_1] \cup [a_2,b_2])$ if there is no further clarification.
In \cite{Faulkner:2017vdd}, it was argued that (\ref{commutator}) can be further upgraded to an expression that relates the bulk and boundary modular flows:
\begin{equation}\label{flow}
    e^{i\boldsymbol{K} \tau} \varphi(x)  e^{-i\boldsymbol{K} \tau}  =   e^{iK \tau} \varphi(x)  e^{-iK \tau}  + \mathcal{O}\left(1/c\right),
\end{equation}
where we work in the Heisenberg picture and think of the modular flow as acting on the operators. We define
\begin{equation}
    \varphi (x,\tau) \equiv e^{iK\tau} \varphi(x) e^{-iK\tau}
\end{equation}
as the modular flowed/evolved operator.
In the following, we will explain why we can use the formula (\ref{flow}) to extract information contained in the island, given that we have the knowledge of the exact modular Hamiltonian $\boldsymbol{K}$.

For concreteness, let's imagine that the bulk vacuum state is perturbed by acting with an unitary evolution $\exp (-i \epsilon \varphi_I (x_0))$, where $\varphi_I(x_0)$ is a local simple operator in the island $[a_1,b_1]$. The perturbed state is still well within the code subspace, thus we can use formula (\ref{flow}) safely. As someone with only access to the radiation, our task is to find out the information about the unitary evolution, such as what the operator $\varphi_I$ is, or its location, etc. Although by the entanglement wedge reconstruction, we know that the operator $\varphi_I$ is encoded in the region $\boldsymbol{[a_2,b_2]}$ in the full non-perturbative description, it must have been encoded in such a complicated way that no simple measurements with few operators and limited accuracy (cannot tell non-perturbative effects) could tell its existence or the value of $x_0$. In other words, if one computes the expectation value of some simple operator $\varphi_B$ in $[a_2,b_2]$, one has
\begin{equation}
    \textrm{tr}[e^{-i\epsilon\varphi_I (x_0)} \boldsymbol{\rho} e^{i\epsilon \varphi_I (x_0)}  \varphi_B] -  \textrm{tr}[\boldsymbol{\rho}   \varphi_B] = 0,
\end{equation}
to all orders in bulk perturbation theory, as demanded by bulk locality. However, suppose we have the knowledge of the microscopic modular Hamiltonian $\boldsymbol{K}$, we could apply a modular flow on the state using $\boldsymbol{K}$, or in the Heisenberg picture, we apply the modular flow to the operator $\varphi_I (x_0)$ as in the left hand side of (\ref{flow}).\footnote{Note that since the bath does not contain dynamical gravity, we do not have to worry about the backreaction on the geometry during this process.} By the equality in (\ref{flow}), to the leading order in bulk perturbation theory, this is equivalent to doing a modular flow on $\varphi_I$ using the bulk modular Hamiltonian $K$. Now importantly, the existence of the island requires that the region $[a_1,b_1]$ and $[a_2,b_2]$ share non-zero mutual information. Another way to say this is that $K$ must be a non-local operator in the sense that it couples operators in $[a_1,b_1]$ and $[a_2,b_2]$. Thus under a modular flow, the bulk operator $\varphi_I (x_0,\tau) \equiv e^{iK\tau} \varphi_I (x_0) e^{-iK\tau}$ will become an operator that is supported on both $[a_1,b_1]$ and $[a_2,b_2]$. Once the operator is supported on $[a_2,b_2]$, one is able to do simple measurements to detect its existence. In other words,
\begin{equation}
\begin{aligned}
    \textrm{tr} [ e^{i\boldsymbol{K}\tau}e^{-i\epsilon\varphi_I (x_0)} \boldsymbol{\rho} e^{i\epsilon \varphi_I (x_0)} e^{-i\boldsymbol{K}\tau}  \varphi_B] &   \approx \textrm{tr} [ e^{iK\tau}e^{-i\epsilon\varphi_I (x_0)} e^{-iK\tau}\boldsymbol{\rho}e^{iK\tau} e^{i\epsilon \varphi_I (x_0)} e^{-iK\tau}  \varphi_B] \\
    & = \textrm{tr} [ e^{-i\epsilon\varphi_I (x_0,\tau)} \boldsymbol{\rho} e^{i\epsilon \varphi_I (x_0,\tau)}   \varphi_B]\\
    & \neq  \textrm{tr}[\boldsymbol{\rho}  \varphi_B].
\end{aligned}
\end{equation}
The approximation on the first line comes from (\ref{flow}), which is expected to be of order $1/c$. The difference between the second line and the third line already arises at order one, and one does not need to carry out measurements with very high accuracy to observe it. More importantly, the difference can be worked out in principle just within the bulk conformal field theory instead of in the microscopic theory, since
\begin{equation}
\begin{aligned}
  & \quad \textrm{tr} [ e^{i\boldsymbol{K}\tau}e^{-i\epsilon\varphi_I (x_0)} \boldsymbol{\rho} e^{i\epsilon \varphi_I (x_0)} e^{-i\boldsymbol{K}\tau}  \varphi_B]  -   \textrm{tr}[\boldsymbol{\rho}   \varphi_B] \\
  & \approx \langle e^{i\epsilon \varphi_I (x_0,\tau)} \varphi_B e^{-i\epsilon \varphi_I (x_0,\tau)} -\varphi_B \rangle_{\textrm{bulk}} + \mathcal{O}(1/c).
\end{aligned}
\end{equation}
Thus by measuring this differences with just simple operators, one can infer about information in the island.

One might complain that computing $e^{iK\tau} \varphi_I (x_0) e^{-iK\tau}$ in the bulk quantum field theory is a taunting task as the bulk modular Hamiltonian of disjoint intervals $[a_1,b_1]\cup [a_2,b_2]$ is generally not known. This suggests that the decoding process should also be complicated in some sense. However, we want to stress that this complication is one in \emph{quantum field theory}, but not one in \emph{quantum gravity}. Indeed, without knowledge of explicit formula for $K$, it is difficult to quantify how much information one can extract from the island in this way. However, in the following sections, we will discuss an example of two dimensional free fermion theory, where the explicit expression of $K$ is known \cite{Casini:2009vk}. There we will show that this way of extracting information has a very simple bulk picture, and is almost perfect in certain cases.

In summary, acting with the exact modular flow $e^{-i\boldsymbol{K}\tau}$ solely on the degrees of freedom in $\boldsymbol{[a_2,b_2]}$, it is as if doing a modular flow $e^{-iK\tau}$ in the combined region $[a_1,b_1]\cup [a_2,b_2]$ in the semiclassical gravity picture. Such a modular flow can carry information from the island $[a_1,b_1]$ to the bath $[a_2,b_2]$. Of course, in order to do it, one must have the knowledge about the microscopic modular Hamiltonian $\boldsymbol{K}$, which is not a simple operator by itself. Our perspective is simply to say that this particular complicated operator provides a general way to extract information from the island, and it has a simple gravity interpretation.

\subsection{Modular flow for $(1 + 1)$-d free massless fermion}\label{sec:formulafermion}

For general field theories, the modular Hamiltonian of disjoint intervals can be quite complicated and is generally unknown, even for the vacuum state. However, for free massless fermion in two dimensions, the modular Hamiltonian of arbitrary disjoint intervals in the vacuum state was derived explicitly in \cite{Casini:2009vk}. This allows us to use the massless fermion as an explicit example to illustrate the idea of pulling out information from the island with modular flow. 
However, we should stress that the idea that was discussed in the previous section applies to general field theories.

\begin{figure}[t!]
\centering

\begin{tikzpicture}[thick,scale = 1.85]

\draw (-2,0) -- (0,2) -- (2,0) -- (0,-2) -- (-2,0);

\draw[cyan] (-1.5,0) -- (-1,0);

\draw[cyan] (-0.5,0) -- (0,0);

\draw[cyan] (0.5,0) node{$......$};

\draw[cyan] (1,0) -- (1.5,0);

\draw[->] (-0.2,-2.2) -- (-2.2,-0.2) ;
\draw (-2.3, -0.1) node{$u_{-}$};

\draw[->] (-2.2,0.2) -- (-0.2,2.2) ;
\draw (-0.2, 2.3) node{$u_{+}$};

\draw[dashed] (-1.5,0) -- (-1.95,-0.45);

\draw  (-1.95-0.15,-0.45-0.1) node{$u_{-,1}^L$};

\draw[dashed] (-1.5,0) -- (-1.95,0.45);

\draw  (-1.95-0.15,0.45+0.2) node{$u_{+,1}^L$};

\draw[dashed] (-1,0) -- (-1.7,-0.7);
\draw  (-1.7-0.15,-0.7-0.1) node{$u_{-,1}^R$};

\draw[dashed] (-1,0) -- (-1.7,0.7);
\draw  (-1.7-0.15,0.7+0.2) node{$u_{+,1}^R$};

\draw[dashed] (-0.5,0) -- (-1.45,-0.95);
\draw  (-1.45-0.15,-0.95-0.1) node{$u_{-,2}^L$};

\draw[dashed] (-0.5,0) -- (-1.45,+0.95);
\draw  (-1.45-0.15,+0.95+0.2) node{$u_{+,2}^L$};

\draw[dashed] (-0,0) -- (-1.2,-1.2);
\draw  (-1.2-0.15,-1.2-0.1) node{$u_{-,2}^R$};

\draw[dashed] (-0,0) -- (-1.2,1.2);
\draw  (-1.2-0.15,1.2+0.2) node{$u_{+,2}^R$};

\draw[dashed] (1,0) -- (-0.7,-1.7);
\draw  (-0.7-0.15,-1.7-0.1) node{$u_{-,n}^L$};

\draw[dashed] (1,0) -- (-0.7,1.7);
\draw  (-0.7-0.15,1.7+0.2) node{$u_{+,n}^L$};

\draw[dashed] (1.5,0) -- (-0.45,-1.95);
\draw  (-0.45-0.15,-1.95-0.1) node{$u_{-,n}^R$};

\draw[dashed] (1.5,0) -- (-0.45,1.95);
\draw  (-0.45-0.15,1.95+0.2) node{$u_{+,n}^R$};

\end{tikzpicture}

\caption{In the discussion of free chiral fermions, it is convenient to think about the intervals in terms of light-cone coordinates.}
\label{fig:lightcone0}
\end{figure}
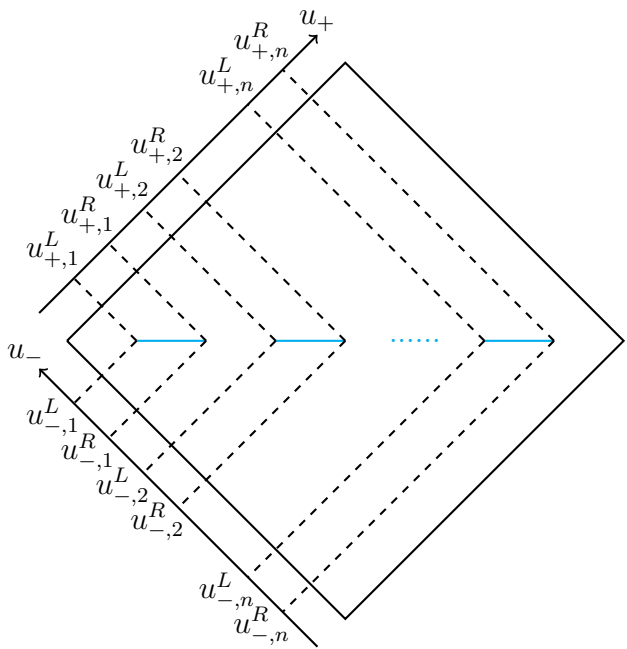

We start with some review of the equations in \cite{Casini:2009vk}. We consider the vacuum state of a free massless Dirac fermion theory in two dimensional flat spacetime. Consider a spacelike region $V$ that contains $n$ disjoint intervals. The modular Hamiltonian of the region $V$ factorizes into two pieces involving fermions with different chiralities:
\begin{equation}\label{factorize}
    K = K_+ + K_-,
\end{equation}
\begin{equation}\label{modularK}
 K_{\pm} = \int_{V_{\pm}} du_{\pm}^{1}\,du_{\pm}^{2}\,\Psi_{\pm}^{\dagger}(u_{\pm}^{1}) H_{\pm}(u_{\pm}^{1},u_{\pm}^{2}) \Psi_{\pm} (u_{\pm}^{2}), \quad u_{\pm} = t\pm x.
\end{equation}
In the formula, $V_{\pm}$ are the projections of the region $V$ onto the light-cone coordinates (see fig. \ref{fig:lightcone0}), which we denote as 
\begin{equation}\label{defregion}
\begin{aligned}
    V_+ & \equiv (u_{+,1}^L, u_{+,1}^R) \cup (u_{+,2}^L, u_{+,2}^R)\cup ... \cup (u_{+,n}^L, u_{+,n}^R), \quad & u_{+,i}^L< u_{+,i}^R, u_{+,i}^{R} < u_{+,i+1}^L,\\
    V_- & \equiv (u_{-,n}^R, u_{-,n}^L) \cup (u_{-,n-1}^R, u_{-,n-1}^L)\cup ... \cup (u_{-,1}^R, u_{-,1}^L),\quad & u_{-,i}^R< u_{-,i}^L, u_{-,i}^{L} < u_{-,i-1}^R.\end{aligned}
\end{equation}

The matrix $H_{\pm}$ in (\ref{modularK}) can be further separated into a local piece and a nonlocal piece which couples the disjoint intervals:
\begin{equation}
    H_{\pm}(x,y) = H_{\pm,\textrm{loc}} (x,y) + H_{\pm,\textrm{noloc}} (x,y).
\end{equation}
The local piece is given by
\begin{equation}\label{local}
    H_{\pm,\textrm{loc}} = \pi i\left(2\left(\frac{d z_{\pm}(x)}{d x}\right)^{-1} \partial_{x}+ \frac{d}{d x}\left(\frac{d z_{\pm} (x)}{d x}\right)^{-1}\right) \delta(x-y),
\end{equation}
where
\begin{equation}\label{z(x)}
    z_{\pm} (x) \equiv \log \left[  - \frac{ \prod_{i=1}^{n} (x - a_{i}^{\pm})}{\prod_{i=1}^{n} (x - b_{i}^{\pm})}    \right].
\end{equation}
Here $a_{i}^{\pm}$ and $b_{i}^{\pm}$ label the positions of the boundary points in the light-cone coordinate. One has $a_{i}^{+} \equiv u_{+,i}^{L}, b_{i}^{+} \equiv u_{+,i}^{R}$, and $a_{i}^{-} \equiv u_{-,(n+1-i)}^{R}, b_{i}^{-} \equiv u_{-,(n+1-i)}^{L}$. With this definition, we can omit the subscript $\pm$ in the following, and we just have to remember that $a_i$ and $b_i$ are defined differently for the left and right moving modes. Note that we always have $a_{i} < b_i, b_{i} < a_{i+1}$.

The function $z(x)$ defined in (\ref{z(x)}) is a monotonic function in each interval $(a_i,b_i)$, which goes from $-\infty$ at $a_i$ to $+\infty$ at $b_i$. By this property of $z(x)$, if we fix a value of $z$, there is one and only one point in each interval which corresponds to it, and we denote it as $x_l (z)$ where $l = 1,...,n$ labels which interval it is in.

\begin{figure}[t!]
 	\center
 	\includegraphics[width=0.6\columnwidth]{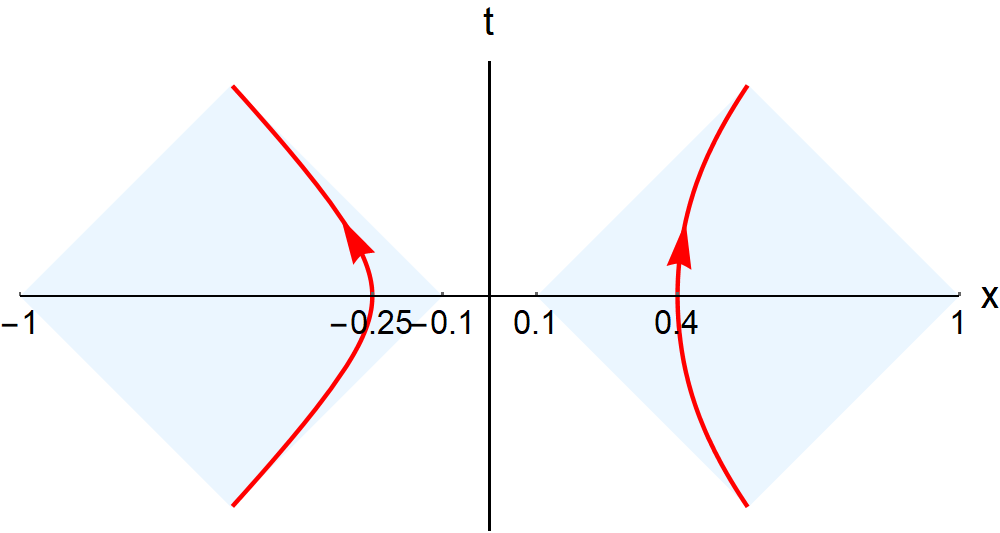}
 	\caption{An example with the region $[-1,-0.1] \cup [0.1,1]$ and $z(0) \approx 0.336$. The causal diamonds of the two intervals are shaded in blue. Under the modular flow, the two fermion operators travel along the red curve while mixing with each other. The positive $\tau$ direction is marked by the arrow. Note that the red curves are \emph{not} generated by the conformal Killing vectors inside each diamond. } 
 	\label{fig:flowexample} 	
 \end{figure}

\begin{figure}[t!]
\centering  

\subfigure[Left-moving operator]  
{  
\includegraphics[width=0.47\columnwidth]{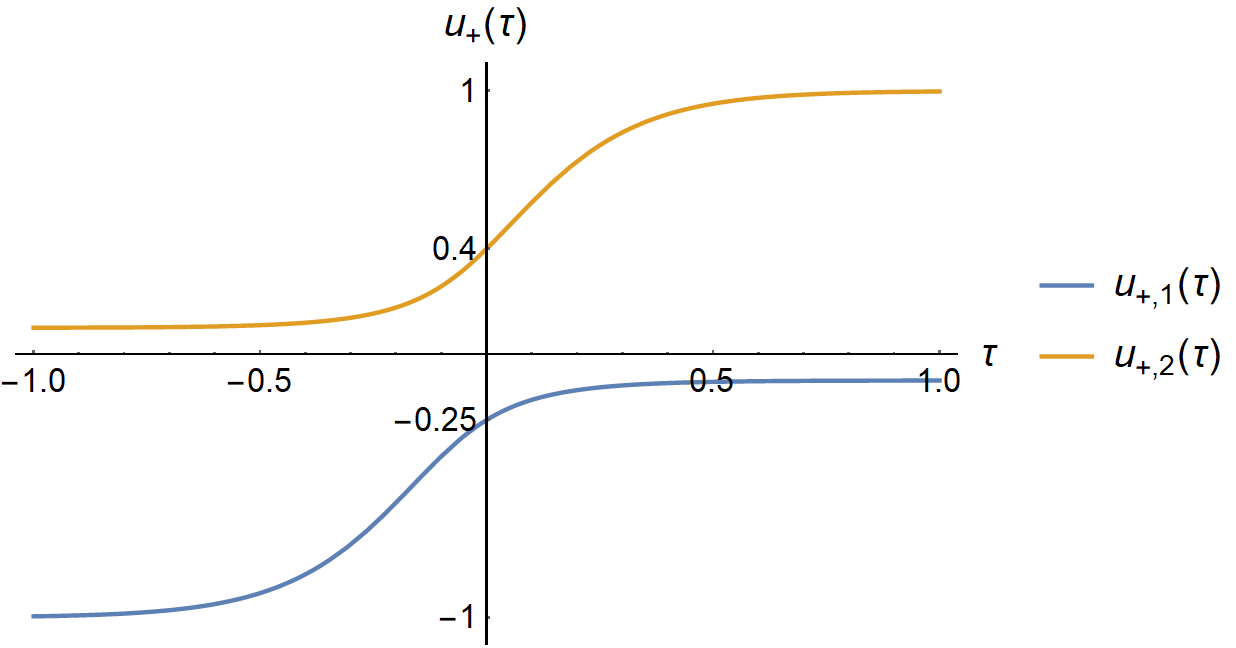}
}  
\subfigure[Right-moving operator]  
{  
\includegraphics[width=0.47\columnwidth]{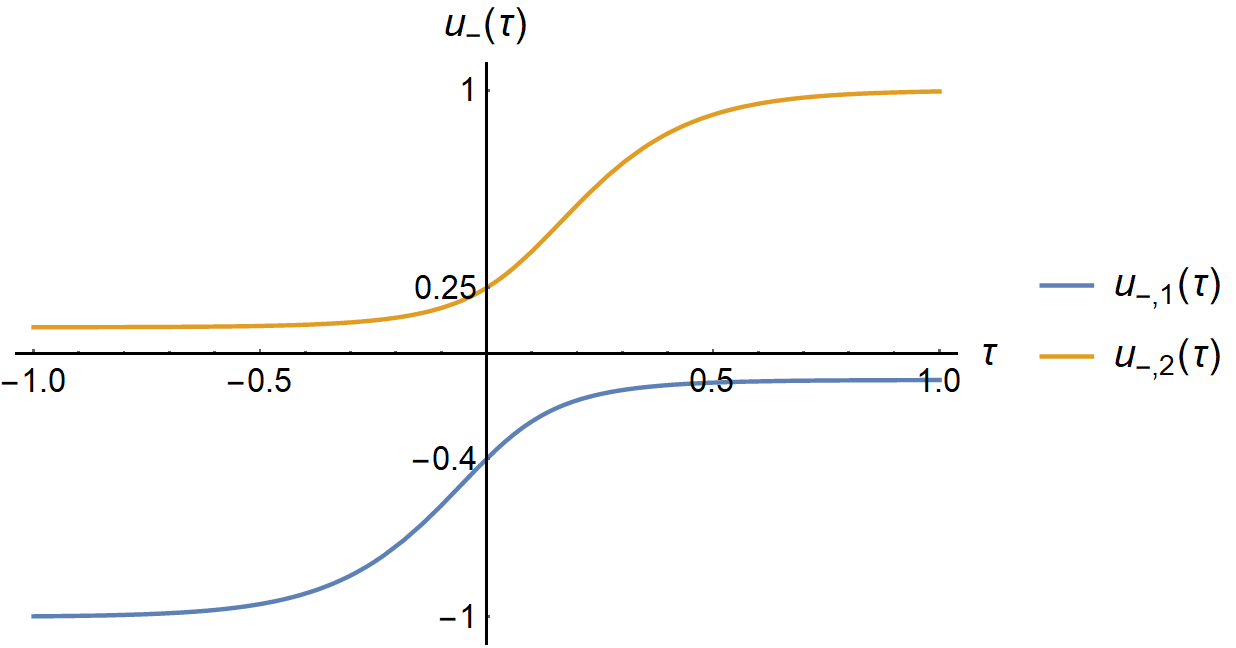}
}
\caption{The trajectories of the left-moving and right-moving operators during the flow in fig. \ref{fig:flowexample}. Note that from the definition of $u_{\pm}$ in (\ref{defregion}), $u_-$ is larger at the left end-point, while $u_+$ is larger at the right end-point, thus  under the modular flow in fig.\ref{fig:flowexample}, the left-moving operator is pushed to the right, while the right-moving operator is pushed to the left. Also note that the regions $1$ and $2$ are also labeled oppositely for operators with different chiralities.}
\label{fig:flowexample2}
\end{figure}

A remarkable property of the non-local piece of the modular Hamiltonian for free massless fermion is that it only couples the points in different intervals with the same $z(x)$. More explicitly, the non-local piece can be written as
\begin{equation}
      H_{\textrm{noloc}} = -2 \pi i \sum_{l  = 1}^{n } \frac{1}{x-y} \left( \frac{dz}{dy}\right)^{-1} \delta (y- x_l(z(x)),\quad x_l (z(x)) \neq x.
\end{equation}
This ``quasi-localness" of the modular Hamiltonian makes the analysis of modular flow particularly simple.  $H_{\textrm{noloc}}$ takes a local operator $\Psi_i (z)$ into $n$ local operators inside the $n$ disjoint intervals with the same $z$. Here the subscript $i$ labels which interval it is in, instead of the chiralities. Under the modular flow, the positions of the $n$ operators flow in a way such that $z(\tau) = z (0) +2\pi \tau$, and the relative weights on the $n$ operators also vary with $\tau$. An example with two intervals $[-1,-0.1] \cup [0.1,1]$ is shown in fig. \ref{fig:flowexample} and fig. \ref{fig:flowexample2}. In the figures, a particular value of $z(0) \approx 0.336$ is chosen. By eqn. (\ref{z(x)}), this sets the positions of the fermion operators at $\tau = 0$ to be $x = -0.25$ and $x=0.4$ in the two diamonds. 

For the case that $V$ is composed of two disjoint intervals $[a_1,b_1]$ and $[a_2,b_2]$, the result of the modular flow is explicitly worked out in \cite{Casini:2009vk}. We absorb an Jacobian factor into the fermion operator by defining
\begin{equation}\label{223}
    \tilde{\Psi}_l (z) \equiv (dx_l / dz)^{\frac{1}{2}}\Psi (x_l (z)).
\end{equation}
Expression (\ref{223}) is just a fermion operator at location $x_l$ determined by parameter $z$. On the other hand, starting from an operator at $x_l(z)$ and apply the modular flow, the resulting operator is denoted as
\begin{equation}\label{224}
    \tilde{\Psi}_l(z,\tau) \equiv e^{iK\tau} \tilde{\Psi}_l(z)e^{-iK\tau}.
\end{equation}
It was showed in \cite{Casini:2009vk} that (\ref{223}) and (\ref{224}) are related in a simple way:
\begin{equation}\label{flowtwo}
    \begin{pmatrix}
        \tilde{\Psi}_1 (z(\tau)) \\
        \tilde{\Psi}_2 (z(\tau))
    \end{pmatrix} 
    =  \begin{pmatrix}
        \cos \theta(\tau) & -\sin \theta (\tau) \\ 
        \sin \theta (\tau) & \cos \theta (\tau)
    \end{pmatrix} 
     \begin{pmatrix}
        \tilde{\Psi}_1 (z(0),\tau) \\
        \tilde{\Psi}_2 (z(0),\tau)
    \end{pmatrix} ,\quad  z(\tau) = z(0) + 2\pi \tau,
\end{equation}
where the function $\theta(\tau)$ is given by:\footnote{The original formula for $\theta(\tau)$ in \cite{Casini:2009vk} was incorrect, as was already noted in \cite{Longo:2009mn}. We thank H. Casini and M. Huerta for verifying this information.}
\begin{equation}
\begin{aligned}
    \theta (\tau)   =  & \arctan \frac{(b_1 + b_2 - a_1 - a_2)x_1 (\tau) + (a_1 a_2 - b_1 b_2)}{\sqrt{(b_1 - a_1) (a_2 - b_1) (b_2 -a_1) (b_2 - a_2)}} \\ 
    & - \arctan \frac{(b_1 + b_2 - a_1 - a_2)x_1 (0) + (a_1 a_2 - b_1 b_2)}{\sqrt{(b_1 - a_1) (a_2 - b_1) (b_2 -a_1) (b_2 - a_2)}}.
\end{aligned}
\end{equation}
\begin{figure}[t!]
 	\center
 	\includegraphics[width=0.6\columnwidth]{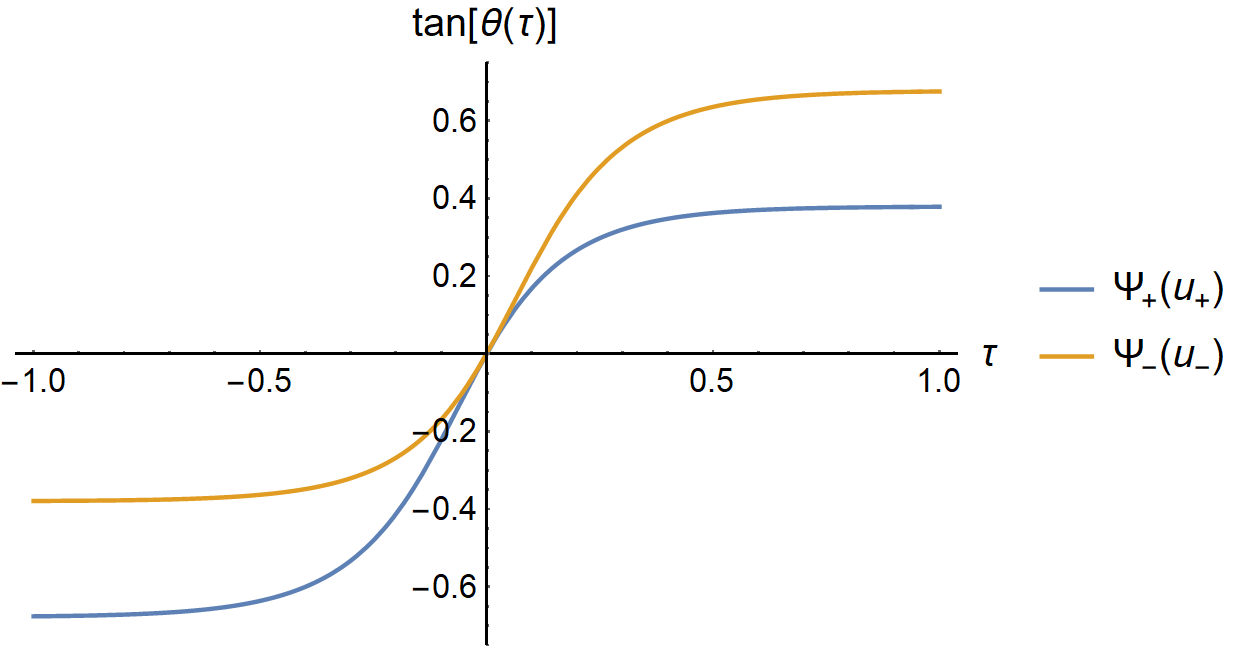}
 	\caption{We plot the function $\tan \theta(\tau)$ for the example shown in fig. \ref{fig:flowexample} and \ref{fig:flowexample2}. The function is not the same for different chiral modes, as one has to substitute in the correct definition of $a_1,b_1,a_2,b_2$ as mentioned after (\ref{z(x)}). The maximal value of $|\tan\theta(\tau)|$ achieved for the example is approximately $0.68$. Since $\tan\theta(\tau)$ does not go to infinity, an operator in one interval is not completely transferred to the other in this example.} 
 	\label{fig:fidelity} 	
 \end{figure}
 In the formula, $x_1(\tau)\in [a_1,b_1]$ is given implicitly by (\ref{z(x)}) and $z(\tau) =z(0)+2\pi \tau$. Since $x_1 (\tau)$ monotonically increases with $\tau$, $\theta (\tau)$ also monotonically increases with $\tau$, starting from $\theta(0)=0$. $\theta(\tau)$ is invariant under the global conformal transformations, and thus we can write it in terms of three independent conformal cross ratios:
\begin{equation}
    \tan \theta (\tau) = \frac{ (\eta_1 (\tau) - \eta_1 (0) ) \sqrt{\eta - 1} }{ (\eta - 1) (\eta_1 (\tau) - 1 )  (\eta_1 (0) - 1 ) + \eta_1 (\tau)\eta_1 (0) },
\end{equation}
where
\begin{equation}
    \eta \equiv  \frac{(a_2 - a_1) (b_2 - b_1)}{(a_2 - b_1)(b_2 -a_1)}, \quad \eta_1 (\tau) \equiv \frac{(b_1 - a_1)( a_2  - x_1 (\tau) ) }{ (b_1 - x_1 (\tau)) (a_2 - a_1)}.
\end{equation}
If we start from an operator in $[a_1,b_1]$, then (\ref{flowtwo}) tells us that after a modular flow with parameter $\tau$, it becomes a linear combination of operators in $[a_1,b_1]$ and $[a_2,b_2]$, i.e.
\begin{equation}
       \tilde{\Psi}_1 (z(0),\tau)  =  \cos \theta(\tau) \tilde{\Psi}_{1} (z(\tau)) + \sin \theta(\tau) \tilde{\Psi}_{2} (z(\tau)) .
\end{equation}
Thus we see that the conformally invariant quantity $\theta(\tau)$ quantifies how ``perfectly" an operator in $[a_1,b_1]$ is being flowed to $[a_2,b_2]$. In general, $\tan\theta(\tau)$ does not go to $\pm \infty$ as $\tau \rightarrow \pm \infty$. In fig. \ref{fig:fidelity}, we plot the function $\tan \theta(\tau)$ for the example shown in fig. \ref{fig:flowexample} and \ref{fig:flowexample2}.

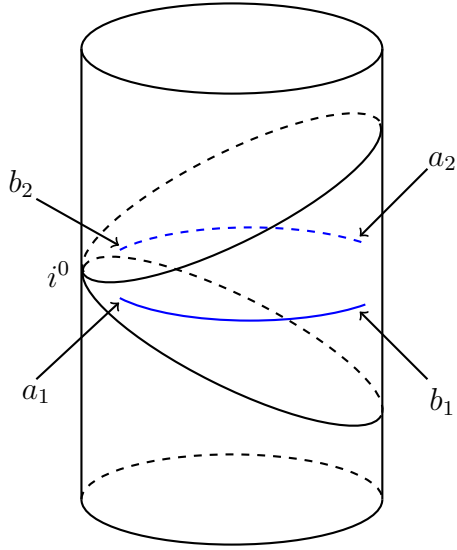
\begin{figure}[t!]
\centering

\begin{tikzpicture}[thick,scale = 2]

\draw[->] (1.3,0.7) -- (0.85,0.25);
\draw (1.4,0.75) node{$a_2$};

\draw[->] (-1.3,0.5) -- (-0.75,0.19);
\draw (-1.4,0.6) node{$b_2$};

\draw (1,-1.5) -- (1,1.5);
\draw (-1,-1.5) -- (-1,1.5);
\draw (-1,1.5) arc (180:0:1 and 0.3);
\draw (-1,1.5) arc (-180:0:1 and 0.3);

\draw[dashed] (-1,-1.5) arc (180:0:1 and 0.3);
\draw (-1,-1.5) arc (-180:0:1 and 0.3) ;

\draw[blue][dashed] (-0.745,0.16) arc (150:40:1 and 0.3) ;

\begin{scope}[shift={(-0.1,0.46)}]
\draw[rotate = 26.56] (-1,0) arc (-180:0:1.105 and 0.3) ;
\end{scope}
\begin{scope}[shift={(-0.09,-0.4)}]
\draw[rotate = -26.56] (-1,0) arc (-180:0:1.105 and 0.3) ;
\end{scope}

\begin{scope}[shift={(-0.09,0.46)}]
\draw[rotate = 26.56,dashed] (-1,0) arc (180:0:1.105 and 0.3) ;
\end{scope}
\begin{scope}[shift={(-0.09,-0.4)}]
\draw[rotate = -26.56,dashed] (-1,0) arc (180:0:1.105 and 0.3) ;
\end{scope}

\draw[blue] (-0.745,-0.16) arc (-150:-40:1 and 0.3) ;

\draw[->] (-1.3,-0.7) -- (-0.75,-0.19);
\draw (-1.3,-0.8) node{$a_1$};

\draw[->] (1.3,-0.7) -- (0.85,-0.25);
\draw (1.4,-0.85) node{$b_1$};


\draw (-1.15,0) node{$i^0$};

\end{tikzpicture}

\caption{If we view the Minkowski spacetime as a patch that is conformally compactified on the surface of a cylinder, then as $a_1$ and $b_2$ are being pushed toward the spatial infinity $i^0$, they are in fact getting closer to each other.}
\label{fig:cylinder}
\end{figure}

We can work out the bounds on the possible value of $\tan\theta(\tau)$. We first consider the limit of $\tau \rightarrow \infty$, where $x_1 (\tau)$ is pushed to $b_1$. In such a limit, $\eta_1 (\infty) \rightarrow \infty$, and
\begin{equation}\label{bound1}
   | \tan \theta(\infty)| = \frac{\sqrt{\eta - 1}}{\eta \eta_1 (0) - (\eta-1)} <  \sqrt{\eta - 1},
\end{equation}
where the optimal value $\sqrt{\eta -1}$ is approached if the initial position of the operator $x_1 (0)$ is close to $a_1$. We can also try to modular evolve the operator in a different direction $\tau \rightarrow -\infty$. In this limit, $x_1 (\tau)\rightarrow a_1$ and $\eta_1 (-\infty) \rightarrow 1$. We find
\begin{equation}\label{bound2}
   | \tan \theta(-\infty)| = \frac{(\eta_1 (0) - 1)\sqrt{\eta - 1}}{\eta_1 (0)} <  \sqrt{\eta - 1},
\end{equation}
where the optimal value $\sqrt{\eta -1}$ is approached if the initial position of the operator $x_1 (0)$ approaches $b_1$. The conclusion is that in general one has
\begin{equation}\label{bound}
       | \tan \theta(\tau)| < \sqrt{\eta-1},
\end{equation}
and with larger value of the cross ratio $\eta$, one has greater room to transfer an operator from one interval to the other. This can be intuitively understood as follows. One example that we have $\eta \rightarrow \infty$ is when we fix $a_1$ and $b_2$, then let $b_1 \rightarrow a_2$. In this limit, physical results should look like as if we had a single interval from $a_1$ to $b_2$. In the single interval case ($b_1 =a_2$), the geometrical modular flow can simply carry an operator from $(a_1,b_1)$ to $(a_2,b_2)$ perfectly. From this intuitive picture, it is also clear that if we fix the initial value $x_1(0)$ of the operator in $(a_1,b_1)$, it would be optimal to flow it in the direction of $\tau >0$, i.e. towards $b_1$. This can be checked by holding $a_1,x_1(0),b_1,b_2$ fixed, and take $a_2\rightarrow b_1$. In this limit, one finds that $\tan\theta(\infty) \rightarrow \infty$, while $\tan\theta(-\infty)\rightarrow 0$.

Another example that $\eta \rightarrow \infty$ is when we consider $b_1$ and $a_2$ fixed, $b_2 = -a_1 = \ell$, while taking $\ell$ to infinity. It seems that $a_1$ and $b_2$ are getting away from each other, but if we view the Minkowski spacetime as a patch that is conformally compactified on the surface of a 2D cylinder (see fig. \ref{fig:cylinder}), $a_1$ and $b_2$ are actually approaching each other and thus improves the extraction. Of course, we could also apply a global conformal transformation which brings the second example to the first one. In this case, if one fixes the positions of $b_1,x_1(0),a_2$, then it is optimal to flow the operator in the negative direction, i.e. towards $a_1$.

\subsection{Pulling out the island}\label{sec:pullout}

Now we apply the discussion of the free massless fermion field to the extremal black hole set-up as discussed in sec. \ref{general}. We could consider a possible scenario where the bulk CFT includes a sector that is or can be approximated by free massless fermion fields, and we are interested in extracting the information in the island that is carried by this fermion field. As we argued previously, doing a modular flow with the microscopic modular Hamiltonian $\boldsymbol{K}$ corresponds to a bulk modular flow in the union $[a_1,b_1] \cup [a_2,b_2]$. In the example of sec. \ref{general}, the bulk quantum fields are simply in the vacuum state, thus apart from $c$-number terms in the modular Hamiltonian that come from the conformal anomaly on the AdS$_2$ space, which do not enter the discussion of the dynamics under the modular flow, the modular Hamiltonian is the same as what we have discussed in sec. \ref{sec:formulafermion}. Thus we can directly apply the results here and write:\footnote{Here we are implicitly looking at the left-moving modes, for which $\tilde{\Psi}_1$ is an operator in the island. For the right-moving modes, one needs to suitably swap the subindices $1$ and $2$, but the physics is the same.}
\begin{equation}
    e^{i\boldsymbol{K}\tau} \tilde{\Psi}_1 (z(0))  e^{-i\boldsymbol{K}\tau}  =   \cos \theta(\tau) \tilde{\Psi}_{1} (z(\tau)) + \sin \theta(\tau) \tilde{\Psi}_{2} (z(\tau)) + \mathcal{O} \left(1/c\right) .
\end{equation}
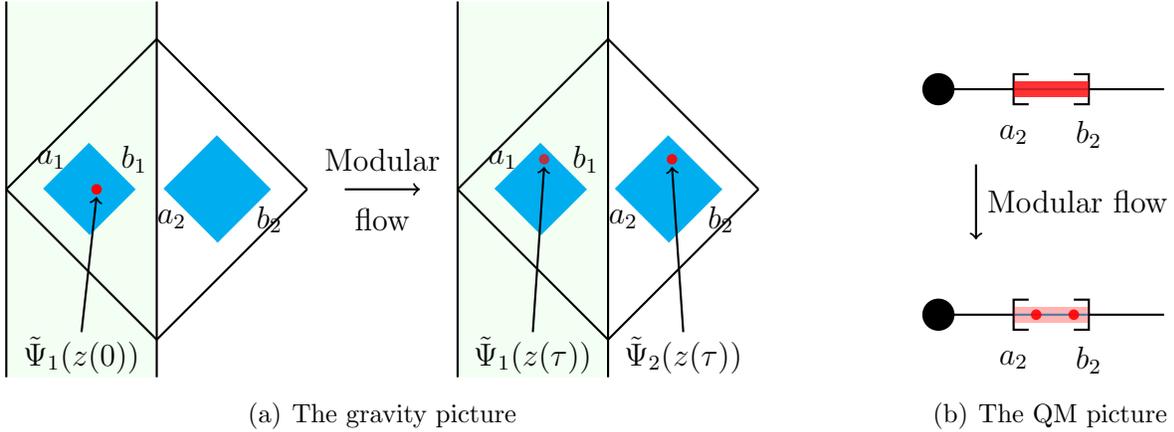
\begin{figure}[t!]
\centering  
\subfigure[The gravity picture]  
{  
 \begin{tikzpicture}[thick,scale = 1]

 \fill[green, opacity = 0.05] (0,-2.5) -- (-2,-2.5) -- (-2,2.5) -- (0,2.5); 
 
 \draw (0,-2.5) -- (0,2.5);
 \draw (-2,-2.5) -- (-2,2.5);
 \draw (-2,0) -- (0,2);
 \draw (-2,0) -- (0,-2);
 \draw (2,0) -- (0,2);
 \draw (2,0) -- (0,-2);

  \draw (0.2,-0.4) node{$a_2$};
  \draw (1.5,-0.4) node{$b_2$};
  \draw (-0.3,0.4) node{$b_1$};
  \draw (-1.4,0.4) node{$a_1$};
\filldraw[cyan] (0.1,0) -- (0.8,0.7)--(1.5,0) -- (0.8,-0.7);
\filldraw[cyan] (-1.5,0) -- (-0.9,0.6)--(-0.3,0) -- (-0.9,-0.6);
 
 \fill[red] (-0.8,0.) circle (2pt); 
  \draw [->] (-1,-1.9) -- (-0.8,-0.1);
  \draw (-1,-2.2) node{$\tilde{\Psi}_1 (z(0))$}; 
 
 \draw[->] (2.5,0) -- (3.5,0);
  \draw (3,0.4) node{Modular};
 \draw (3,-0.4) node{flow};
 
  \begin{scope}[shift={(6,0)}]
  
 \fill[green, opacity = 0.05] (0,-2.5) -- (-2,-2.5) -- (-2,2.5) -- (0,2.5); 
 
   \draw (0,-2.5) -- (0,2.5);
 \draw (-2,-2.5) -- (-2,2.5);
 \draw (-2,0) -- (0,2);
 \draw (-2,0) -- (0,-2);
 \draw (2,0) -- (0,2);
 \draw (2,0) -- (0,-2);

  \draw (0.2,-0.4) node{$a_2$};
  \draw (1.5,-0.4) node{$b_2$};
  \draw (-0.3,0.4) node{$b_1$};
  \draw (-1.4,0.4) node{$a_1$};
\filldraw[cyan] (0.1,0) -- (0.8,0.7)--(1.5,0) -- (0.8,-0.7);
\filldraw[cyan] (-1.5,0) -- (-0.9,0.6)--(-0.3,0) -- (-0.9,-0.6);

 \fill[red,opacity=0.7] (-0.85,0.4) circle (2pt); 
  \draw [->] (-1,-1.9) -- (-0.85,0.3);
  \draw (-1,-2.2) node{$\tilde{\Psi}_1 (z(\tau))$};
  
 \fill[red,opacity=0.9] (0.85,0.4) circle (2pt); 
  \draw [->] (1,-1.9) -- (0.85,0.3);
  \draw (1,-2.2) node{$\tilde{\Psi}_2 (z(\tau))$};

  \end{scope}
\end{tikzpicture}

}  
\hspace{0.1\textwidth}
\subfigure[The QM picture]  
{  
 \begin{tikzpicture}[thick,scale = 1]
 \filldraw[draw = black] (6,0) circle (0.2);
 \draw (6,0) -- (7,0);
 \draw[cyan, thick] (7,0) -- (8,0);
 \draw (8,0) -- (9,0);
  \draw[thick] (7.2,-0.2)--(7,-0.2) -- (7,0.2) -- (7.2,0.2);
  \draw[thick] (7.8,-0.2)--(8,-0.2) -- (8,0.2) -- (7.8,0.2);
  
  \draw (7,-0.6) node{$a_2$};
  \draw (8,-0.6) node{$b_2$};
  
  \fill[red,opacity=0.8] (7.01,-0.1) -- (7.01,0.1) -- (7.99,0.1) -- (7.99,-0.1) ;

  \begin{scope}[shift={(0,-3.)}]
  \filldraw[draw = black] (6,0) circle (0.2);
 \draw (6,0) -- (7,0);
 \draw[cyan, thick] (7,0) -- (8,0);
 \draw (8,0) -- (9,0);
  \draw[thick] (7.2,-0.2)--(7,-0.2) -- (7,0.2) -- (7.2,0.2);
  \draw[thick] (7.8,-0.2)--(8,-0.2) -- (8,0.2) -- (7.8,0.2);
  
  \draw (7,-0.6) node{$a_2$};
  \draw (8,-0.6) node{$b_2$};
  
  \fill[red,opacity=0.3] (7.01,-0.1) -- (7.01,0.1) -- (7.99,0.1) -- (7.99,-0.1) ; 
    \fill[red,opacity=0.9] (7.8,0) circle (2pt); 
      \fill[red,opacity=0.9] (7.3,0) circle (2pt);

  \end{scope}
  \draw[->] (6.5,-1) -- node[right]{Modular flow}(6.5,-2);

\end{tikzpicture}
}

\caption{(a) The gravity picture of the modular flow: a fermion operator in the island is partially pulled out to the bath. (b) The dual quantum mechanical picture: two simple operators (represented by the red dots) are partially pulled out from a complicated operator $\tilde{\Psi}_1 (z(0))$ (represented by the red cloud). Note there are two dots because we have two different chiral components.}
\label{fig:pulling}
\end{figure}
In fig. \ref{fig:pulling}, we illustrate this formula with pictures. In fig. \ref{fig:pulling}(a), we present the gravity picture: an operator inside the island is partially pulled out to the bath by the modular flow. In fig. \ref{fig:pulling}(b), we also draw an illustration of the quantum mechanical interpretation. What we are doing is to pull out some simple operators from the initially complicated operator ``cloud".

With the choice of parameters in (\ref{parameter}), (\ref{parameter2}), we have
\begin{equation}
    \eta \approx  \frac{b_2}{ 3 \frac{\phi_r}{c}} \gg 1,
\end{equation}
thus by (\ref{bound}) we see that one has the ability of pulling out a large amount of information from the island.\footnote{We've used $\theta(\tau)$ as a quantifier for the amount of information one can extract, in the sense that if $\theta$ stays zero, one could extract no information and if $\theta$ goes to $\pm\pi/2$, the extraction would be perfect. It would be interesting to find a more precise quantifier for intermediate values of $\theta$.}

One might worry that to achieve a large value of $|\tan \theta(\tau)|$, one has to do a modular flow with very large $\tau$, which might invalidates (\ref{flow}) if $\tau$ is comparable to $c$. However, it is easy to see that this does not happen. This is because $\tau$ enters the expression of $x_1(\tau)$ only in the form of $\exp(2\pi\tau)$ and thus the time scale at which $|\tan \theta(\tau)|$ becomes exponentially close to its asymptotic value can only depend on the cross ratios in a logarithmic way and will not reach the scale $c$. 

Above, we described an operational way of pulling out an operator from the island. The same physics can also be phrased in the language of entanglement wedge reconstruction. In \cite{Jafferis:2015del,Faulkner:2017vdd}, it was proposed that modular flow serves as a natural way to generalize the HKLL reconstruction \cite{Banks:1998dd,Hamilton:2006az} to the operators in the entanglement wedge. The idea is that an operator in the entanglement wedge can be reconstructed on the boundary using the modular evolved boundary operators. In our current set up, the formula has a very simple form. From (\ref{flowtwo}) one gets
\begin{equation}
    \tilde{\Psi}_1 (z(0)) = \frac{1}{\sin \theta (\tau)} e^{-i\boldsymbol{K}\tau}\tilde{\Psi}_2 (z(\tau)) e^{i\boldsymbol{K}\tau}- \frac{1}{\tan \theta (\tau)} \tilde{\Psi}_{2} (z(0)) + \mathcal{O}\left(1/c\right),
\end{equation}
that is one can reconstruct an operator in the island using operators in the bath region $[a_2,b_2]$ and its microscopic modular Hamiltonian $\boldsymbol{K}$.

\section{Nonzero temperature black hole coupled to a bath}\label{twoside}

\begin{figure}[t!]
\centering

\begin{tikzpicture}[thick,scale = 1.]

\fill[green, opacity = 0.05] (0,-4) -- (-4,-4) -- (-4,4) -- (0,4); 

\filldraw[cyan] (0.588,0.7666) -- (0.9107,1.0893) -- (2,0) -- (0.588+2-0.9107,0.7666-1.0893);

 \fill[black,opacity=1] (0.588,0.7666) circle (1pt); 

\filldraw[cyan] (-1.414,0.449) -- (-2,1.03528) -- (-2+1.414-2,0.449) -- (-2,0.449+0.449-1.03528);

 \fill[black,opacity=1] (-1.414,0.449) circle (1pt);

 \fill[black,opacity=1] (-2+1.414-2,0.449)  circle (1pt); 

\draw (0.5,0.5) node{$u_{\pm,B_R}$};

 \draw (0,-4) -- (0,4);
 \draw (-2,0) -- (0,2);
 \draw (-2,0) -- (0,-2);
 \draw (2,0) -- (0,2);
 \draw (2,0) -- (0,-2);

\draw[dashed] (0,-2) -- (-2,-4);
\draw[dashed] (-2,-4) -- (-4,-2);

\draw[dashed] (0,2) -- (-2,4);
\draw[dashed] (-2,4) -- (-4,2);
 
\draw (-2,-2)  node{AdS$_2$};
 \draw[->] (2.5,1) -- (3.5,1) node[right]{$x$};
 \draw[->] (2.5,1) -- (2.5,2) node[above]{$t$};
 \draw[->] (2.5,1) -- (3.2,1.7) node[right]{$u_+$};
 \draw[->] (2.5,1) -- (1.8,1.7) node[left]{$u_-$};

 \draw[dotted] (-2,0) arc (135:45:2.828 and 2.828);

\begin{scope}[shift = {(-4,0)}]

\filldraw[cyan] (-0.588,0.7666) -- (-0.9107,1.0893) -- (-2,0) -- (-0.588-2+0.9107,0.7666-1.0893);

\draw (-0.4,0.5) node{$u_{\pm,B_L}$}; 

 \fill[black,opacity=1] (-0.588,0.7666) circle (1pt); 

 \draw (0,-4) -- (0,4);
 \draw (-2,0) -- (0,2);
 \draw (-2,0) -- (0,-2);
 \draw (2,0) -- (0,2);
 \draw (2,0) -- (0,-2);

\draw[dotted] (-2,0) arc (135:45:2.828 and 2.828);
  
\end{scope}

\draw[->] (1,-2) -- (0.5,-1);
\draw (1,-2.2) node{Bath$_R$};
 
\draw[->] (-4-1,-2) -- (-4-0.5,-1);
\draw (-4-1,-2.2) node{Bath$_L$};


\draw (-1.1,0.2) node{$u_{\pm,I_R}$}; 
 \draw (-2.7,0.2) node{$u_{\pm,I_L}$}; 

\end{tikzpicture}

\caption{We have two AdS black holes coupled to baths. If we consider the entanglement wedge of the bath region shown in the figure, it includes an island in the bulk at late time.}
\label{fig:penrose2}
\end{figure}
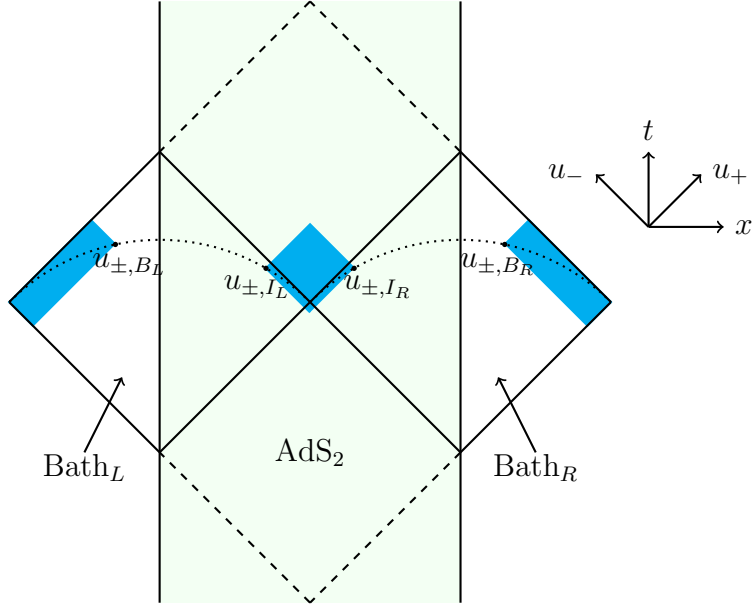

In this section, we generalize the discussion in the previous section to the set up of nonzero temperature black hole case discussed in \cite{Almheiri:2019yqk}. More precisely, we still consider two dimensional JT gravity coupled to a CFT with central charge $c$, but the solution involves two black holes coupled to two baths. The two black hole-bath pairs are prepared in a thermofield double state at $t=0$. The metric inside the diamond completed by the dashed lines in fig. \ref{fig:penrose2} is conformally equivalent to the flat space metric $ds^2 = -dt^2 + dx^2$. We put the origin $(x,t) = (0,0)$ at the bifurcation surface. The conformal fields are again in the Minkowski vacuum, but this time in a larger diamond as in fig. \ref{fig:penrose2}.

We now consider the entanglement wedge of a bath region which goes from the two points $(u_{+,B_{L}},u_{-,B_L})$ and $(u_{+,B_{R}},u_{-,B_R})$ to the spatial infinity. The subscripts $B_L$ and $B_R$ refer to whether the point is in the left bath or the right bath. The coordinates of the two points are given explicitly by
\begin{equation}
    u_{\pm, B_L} = \mp \exp \left( \mp \frac{2\pi (t_R \mp b)}{\beta}  \right), \quad  u_{\pm, B_R} = \pm \exp \left( \pm \frac{2\pi (t_R \pm b)}{\beta}  \right),
\end{equation}
where $\beta$ is the inverse temperature of the black hole, $t_R$ is a parameter which corresponds to the Rindler time in the right wedge, and $b$ is a fixed positive constant. In fig. \ref{fig:penrose2}, we marked the domains of dependence for these two regions in blue.
The result of \cite{Almheiri:2019yqk} is that, the entanglement wedge of the bath region discussed above will include an island located inside the AdS$_2$ region, after the Page time $t_{R,\textrm{Page}} \sim \beta S_{\textrm{BH}}/c$. The left and right boundary points of the island are located at
\begin{equation}
    u_{\pm, I_L} = \mp \exp \left( \mp \frac{2\pi (t_R \mp a)}{\beta}  \right), \quad  u_{\pm, I_R} = \pm \exp \left( \pm \frac{2\pi (t_R \pm a)}{\beta}  \right),
\end{equation}
where $a$ is negative and is related to $b$ via
\begin{equation}
    a \approx -\left( b + \frac{\beta}{2\pi} \log \left(\frac{24\pi \phi_r}{c\beta}\right) \right), \quad \textrm{for} \quad \frac{\phi_r}{c\beta} \gg 1.
\end{equation}
Similar to what we have discussed in sec. \ref{general}, when the entanglement wedge of the bath region contains the island, a modular flow using the microscopic modular Hamiltonian on the bath region corresponds to a bulk modular flow on the union of the bath region and the island, which can bring out the information hidden in the island. 

We could also consider a scenario that the bulk CFT contains a free massless fermion field, and ask how the information extraction works. One may wonder whether the formulas in \ref{sec:formulafermion} still apply, as here the four boundary points $u_{\pm, B_L}, u_{\pm, I_L}, u_{\pm, I_R}, u_{\pm, B_R}$ do not lie on the constant Minkowski time slice. The answer is yes. The reason is that there is no correlation in the vacuum state between fermion operators with different chiralities, and the modular Hamiltonian always factorizes into different chiral pieces as in (\ref{factorize}), no matter the arrangement of the boundary points. Thus we can simply project the four boundary points onto the light-cone coordinates, and study the modular flow for left-moving and right-moving modes separately. Another concern might be that it seems we have three disjoint intervals in this case, one from $u_{\pm, B_L}$ to spatial infinity, one from $u_{\pm, B_R}$ to spatial infinity and the third one is the island. However, we should really view the first two intervals as joined together through the spatial infinity. For example, we could do a conformal transformation which brings the spatial infinity to finite distance, and we literally have two intervals.\footnote{This way would produce the correct entropy for the region, while if one start from three intervals and then push two end points to infinity, one do not get the correct result.} Another viewpoint is that if we view the Minkowski space as a patch that is conformally compactified on a cylinder as in fig. \ref{fig:cylinder}, we would have two intervals instead of three. Thus we see that the discussion in sec. \ref{sec:formulafermion} is still applicable here.

\begin{figure}[t!]
\centering  

\begin{tikzpicture}[thick,scale = 1]

 \draw[cyan] (-3.5,0) -- (-2,0);
 
 \draw (-2.2,-0.2) -- (-2,-0.2) -- (-2,0.2) -- (-2.2,0.2);
 
 \draw (-2,0) -- (-1.5,0);
 
  \draw (-1.3,-0.2) -- (-1.5,-0.2) -- (-1.5,0.2) -- (-1.3,0.2);

 \draw[cyan] (-1.5,0) -- (-0.5,0);
 
  \draw (-0.7,-0.2) -- (-0.5,-0.2) -- (-0.5,0.2) -- (-0.7,0.2);
  
 \draw (-0.5,0) -- (2,0);
  
   \draw (2.2,-0.2) -- (2,-0.2) -- (2,0.2) -- (2.2,0.2);
 
 \draw [cyan] (2,0) -- (3,0);
 
 \draw (-2.2,-0.5) node{\tiny{$u_{+,B_L}$}};
  \draw (-1.5,0.5) node{\tiny{$u_{+,I_L}$}};
  \draw (-0.5,0.5) node{\tiny{$u_{+,I_R}$}};
  \draw (2,-0.5) node{\tiny{$u_{+,B_R}$}}; 
  
  \draw [->] (0,-0.5) -- (0,-1.5);
  \draw (0.9,-1) node{\tiny{$u_+ \rightarrow f(u_+)$}};
 
\begin{scope}[shift = {(0,-2.5)}]

 \draw (-3.5,0) -- (-2.5,0);
 
    \draw (-2.3,-0.2) -- (-2.5,-0.2) -- (-2.5,0.2) -- (-2.3,0.2); 
 
 \draw[cyan] (-2.5,0) -- (-0.5,0);
  
   \draw (-0.7,-0.2) -- (-0.5,-0.2) -- (-0.5,0.2) -- (-0.7,0.2); 
 
 \draw (-0.5,0) -- (-0,0);
 
    \draw (0.2,-0.2) -- (-0.,-0.2) -- (0,0.2) -- (0.2,0.2); 
 
 \draw[cyan] (-0,0) -- (1.5,0);
 
 \draw (1.5,0) -- (3,0);

   \draw (1.3,-0.2) -- (1.5,-0.2) -- (1.5,0.2) -- (1.3,0.2);

 \draw (-2.5,-0.5) node{\tiny{$f(u_{+,I_R})$}};
  \draw (-0.5,-0.5) node{\tiny{$f(u_{+,I_L})$}};
  \draw (0,0.5) node{\tiny{$f(u_{+,B_L})$}};
  \draw (1.5,0.5) node{\tiny{$f(u_{+,B_R})$}}; 
  
\end{scope} 
\end{tikzpicture}

\caption{We can apply a M\"{o}bius transformation such that the regions are arranged in the way we discussed in sec. \ref{sec:formulafermion}. }
\label{fig:lightcone}
\end{figure}
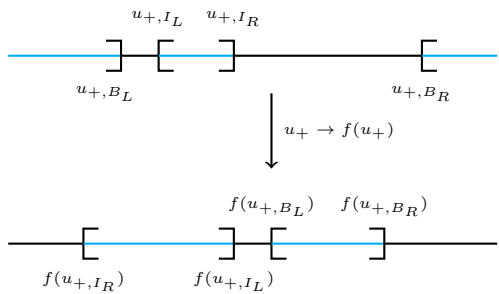

Let's first discuss how the bulk modular flow acts on the left-moving modes. Projecting the boundary points to the $u_+$ axis, we get the top figure in fig. \ref{fig:lightcone}. Under a M\"{o}bius transformation $u_+ \rightarrow f(u_+)$, we can rearrange the points as in the bottom figure in fig. \ref{fig:lightcone}. The explicit form of the M\"{o}bius transformation is not crucial here. The important feature is that we have
\begin{equation}
  u_{+ ,I_L} - u_{+,B_L} \propto  \exp \left(-\frac{2\pi t_{R}}{\beta}\right),
\end{equation}
thus $u_{+,I_L}$ and $u_{+,B_L}$ approach each other exponentially at late time. 
After the transformation, $f(u_{+,B_L})$ and $f(u_{+,I_L})$ also approach exponentially. As we mentioned in sec. \ref{sec:formulafermion}, as two boundary points approach each other, the extraction of an operator becomes optimal. For an left-moving operator with some fixed $u_{+}$ in the island, a modular flow in the negative $\tau$ direction which pushes it to $u_{+,I_L}$ will be able to almost pull it out to the left bath region. Of course, if this left-moving fermion is thrown into the black hole from the right boundary, one needs to wait for a scrambling time before it is included in the island to be pulled out \cite{Almheiri:2019yqk}. Thus our proposal here serves as an example of the Hayden-Preskill decoding process \cite{Hayden:2007cs}. Similar arguments apply for right-going modes, where one also finds that a modular flow in the negative $\tau$ direction can almost pull out the operator from the island to the right bath.

Note that as $t_R$ increases, the island covers a larger part of the black hole interior. Thus the proposed way is able to pull out information from the black hole interior, as far as the semiclassical bulk picture still holds. What one needs is of course, the knowledge of the microscopic density matrix of the bath region.

\section{Conclusion and discussion}

In this paper we've discussed a way to pull out the information from the island to the bath, by acting only on the degrees of freedom in the bath, but with the microscopic modular Hamiltonian. What we did is to apply the conventional wisdom of subregion-subregion duality in holographic theories to the new set-ups involving a gravitational system coupled to a bath, where islands arise. For the examples that the bulk fields include a free massless fermion field, we find that one can almost pull out an operator from the island in both the extremal and the non-extremal black hole cases. Although we've only discussed the situations that the black hole and the bath are in equilibrium, the same idea also applies to more realistic black hole evaporation set-up, such as the ones in \cite{Penington:2019npb,Almheiri:2019psf}, or higher dimensional situations \cite{Almheiri:2019psy}.

In the bulk semiclassical picture, the reason that an island exists is that there is enough mutual information between the island and the bath region, to win over the cost of including extra area contributions as in (\ref{newentropyrule}). Our proposal exactly relies on and utilizes the nonzero mutual information and its resulting nonlocalness in the bulk modular Hamiltonian. In this sense, the resource needed to extract information from the island is simply contained in the bulk quantum fields. At face value, it appears that gravity, especially those non-perturbative effects discussed in \cite{Almheiri:2019qdq,Penington:2019kki} did not play an important role in the extraction process. However, this is not true, as the magic of gravity is to give rise to the formula (\ref{JLMS}), i.e., to encode the information of the island into the microscopic modular Hamiltonian.

Although we put a lot of focus on the example of free fermion field, we should stress that the general idea of information extraction with modular flow should work for general field theories, although the details will be different. We discussed the free fermion field just because we are able to derive analytical results in this case, which helps us demonstrate the idea. In \cite{Arias:2018tmw}, the modular Hamiltonian for free chiral bosons in two intervals was derived, which does not have the ``quasi-local" property as the free fermion case. However, one might expect that when the two intervals approach each other, such as in the example of sec. \ref{twoside}, one can still extract information from one to the other nicely. It would be interesting to check whether this is true.

In \cite{Almheiri:2019yqk}, it is proposed that in certain situation, one can also extract part of the information from the island using the Gao-Jafferis-Wall protocol \cite{Gao:2016bin} (see also \cite{Maldacena:2017axo}). However, the Gao-Jafferis-Wall protocol described in \cite{Almheiri:2019yqk} requires that the island is close to a black hole horizon region that one has access to. This is not true in general, for example in the cases discussed in this paper. The advantage of using modular flow is that it applies to general cases that an island is included in the entanglement wedge.

One might worry whether our proposal violates causality in some potential ways, as we can extract information from a spacelike separated island, by acting only on the bath region. In the dual quantum mechanical description, there is no trouble, since the information we want to extract was already contained in the bath region in some complicated way. In the semiclassical gravity description, there is again no violation of causality, since we are acting with a bulk modular flow which directly couples the island and the bath. It's only questionable when we start to think about the exact non-perturbative gravity description. Since we do not have a complete understanding of it, we cannot draw any conclusions here. It might be that in the full non-perturbative gravity description, the island is connected to the bath in some new way, in the spirit of \cite{Almheiri:2019hni,Maldacena:2013xja}. The causality, with the correct definition, might still be preserved there.

\paragraph{Acknowledgement}
We would like to thank Juan Maldacena for helpful discussions and valuable advice on the manuscript. We also want to thank Ho Tat Lam, Henry Lin and Pengfei Zhang for intriguing conversations. Y.C. is supported by a Centennial Fellowship from Princeton University.

\bibliographystyle{JHEP}
\nocite{*}
\bibliography{cite}

\end{document}